\documentclass[12pt]{article}
\usepackage{times}

\usepackage[scaled=0.92]{helvet}	% set Helvetica as the sans-serif font

\usepackage{graphicx}
\usepackage{color}

\renewcommand\sec[1]{\vspace{0.05in}\noindent{{\large\bf{#1}}}
\addtocounter{section}{1}\setcounter{subsection}{0}
}

\renewcommand\b{\begin{equation}}
\newcommand\e{\end{equation}}

\newcommand{\eq}{\!=\!}
\newcommand{\gt}{\!>\!}
\newcommand{\lt}{\!<\!}

\newcommand{\chaos}{{\mbox{\scriptsize chaos}}}
\newcommand{\osc}{{\mbox{\scriptsize osc}}}
\newcommand{\sgn}{{\mbox{sgn}}}

\begin{document}

%Title Page
\thispagestyle{empty}
\vspace*{0.5in}
\begin{center}
\begin{Large}
{\bf
Stimulus-Dependent Suppression of Chaos
\\
\vspace*{0.05in}
in Recurrent Neural Networks\\
}
\end{Large}
\vspace*{0.2in}  
{\bf Kanaka Rajan}\\
\vspace*{0.05in}
Lewis-Sigler Institute for Integrative Genomics\\
Icahn 262, Princeton University\\
Princeton NJ 08544 USA\\
\vspace*{0.1in}
{\bf L.F. Abbott}\\
\vspace*{0.05in}
Center for Neurobiology and Behavior\\
Department of Physiology and Cellular Biophysics\\
Columbia University College of Physicians and Surgeons\\
New York NY 10032-2695 USA\\
\vspace*{0.05in}
and\\
\vspace*{0.05in}
{\bf Haim Sompolinsky}\\
\vspace*{0.05in}
Racah Institute of Physics\\
Interdisciplinary Center for Neural Computation\\
Hebrew University\\
Jerusalem Israel

\vspace*{0.5in}

{\bf Abstract}

\end{center}

Neuronal activity arises from an interaction between ongoing firing generated spontaneously by neural circuits and responses driven by external stimuli.  Using mean-field analysis, we ask how a neural network that intrinsically generates chaotic patterns of activity can remain sensitive to extrinsic input.  We find that inputs not only drive network responses, they also actively suppress ongoing activity, ultimately leading to a phase transition in which chaos is completely eliminated.  The critical input intensity at the phase transition is a non-monotonic function of stimulus frequency, revealing a ``resonant" frequency at which the input is most effective at suppressing chaos even though the power spectrum of the spontaneous activity peaks at zero and falls exponentially.  A prediction of our analysis is that the variance of neural responses should be most strongly suppressed at frequencies matching the range over which many sensory systems operate. 

\newpage
Circuits of the central nervous system exhibit temporally irregular ongoing activity that is not directly related to sensory or behavioral events. The fact that this spontaneous activity is not suppressed by averaging over the large number of synaptic inputs to each neuron $[1]$ suggests that chaotic network dynamics may represent a substantial local source of fluctuating activity in cortical and subcortical circuits.  Previous modeling studies have shown that nonlinear random network models with strong recurrent excitatory and inhibitory connections generically exhibit chaotic dynamics $[2,3,4]$.  In this work, we ask how intrinsically generated fluctuating activity affects neuronal responses to external stimuli.  The nonlinear effects of oscillatory drive, including frequency dependence and phase locking, have been well explored in low-dimensional chaotic dynamical systems (see e.g. $[5,6,7,8,9]$).  However, relatively few studies have explored entrainment of extended, high-dimensional spatiotemporal chaotic systems by external forcing (see e.g. $[10,11,12,13,14]$). Here, we explore the locking of large chaotic neuronal networks to external stimuli and study how it depends on stimulus amplitude and frequency.  

We study phenomenological firing-rate network models representing neurons in a localized circuit that are coupled by relatively strong excitatory and inhibitory connections randomly distributed in the network.  Specifically, we consider a network of $N$ interconnected neurons, each described by an activation variable $x_i$ for $i\eq 1, 2, \ldots N$, satisfying
\b
\frac{dx_i}{dt} = -x_i + \sum_{j=1}^N J_{ij}\phi(x_j) + H_i  \, ,
\label{netEq}
\e
with $\phi(x_i)$, which is a saturating monotonic function of the total synaptic input $x_i$, representing a normalized firing rate relative to a fixed background rate, $r_0$. Here we choose  
\b 
\phi(x) = \left\{  \begin{array}{ll}
r_0 \tanh\left(x/r_0\right) & {\rm for}\,\,\, x \leq 0\\
(2-r_0)\tanh\left(x/(2-r_0)\right) & {\rm for} \,\,\, x > 0\, ,
\end{array}
\right.
\label{phiDef}
\e
so that the normalized firing rate varies from zero to $2$.  For $r_0\eq 1$, we recover the often-used $\tanh$ function, but we use a smaller value of $r_0 \eq 0.1$, which is more biologically reasonable $[15]$.  The time variable in Eq.\ \ref{netEq} is defined in units of the single-neuron time constant, $\tau_r\eq 10$ ms. Each element of the network connectivity matrix $J$ is chosen randomly and independently $[16]$ from a Gaussian distribution with zero mean and variance $g^2/N$,  where the gain $g$ acts as the control parameter of the network.  The external input term is set to $H_i  \eq I\cos(\omega t + \theta_i)$, with the phase $\theta_i$ chosen randomly and independently for each neuron from a uniform distribution between $0$ and $2\pi$.  This corresponds to situations in which the oscillatory input does not introduce global temporal phase coherence, which occurs, for example, for a population of neurons with a broad range of preferred spatiotemporal phases. 

To characterize the activity of the network, we make extensive use of the autocorrelation function of each neuronal rate averaged across all the units of the network,
\b
C(\tau) = \frac{1}{N}\sum_{i=1}^N \big\langle\phi(x_i(t))\phi(x_i(t+\tau))\big\rangle \, ,
\label{CDef}
\e
where the angle brackets denote a time average.  $C(0)$ is related to the total variance in the fluctuations of the firing rates of the network units, whereas $C(\tau)$ for non-zero $\tau$ provides information about the temporal structure of network activity. 

Previous work $[2]$ has shown that, in the limit $N\!\rightarrow\!\infty$ with no input ($I\eq 0$), this model displays only two types of activity: a trivial fixed point with all $x\eq 0$ when $g\lt 1$ and chaos when $g\gt 1$.  The spontaneously chaotic state is characterized by highly irregular firing rates (Fig.\ 1a), a decaying average autocorrelation function (Fig.\ 1d),  and a continuous power spectrum (Fig.\ 1g).  Note that the fluctuations in Fig.\ 1a are considerably slower than the 10 ms time constant of the model.  The associated average autocorrelation function decays to zero  as $\tau$ increases (Fig.\ 1d) implying that the temporal fluctuations of the spontaneous activity are uncorrelated over large time intervals, a characteristic of the chaotic state. The power spectrum decays from a peak at zero (Fig.\ 1g) and, although it is broad, the power at high frequency is exponentially suppressed.  Strong suppression of high-frequency fluctuations is another characteristic of the chaotic state in these networks. By comparison, the power spectrum of a non-chaotic network responding to a white-noise input falls off only as a power law at high frequencies.  

\begin{figure}
\includegraphics[width=\textwidth]{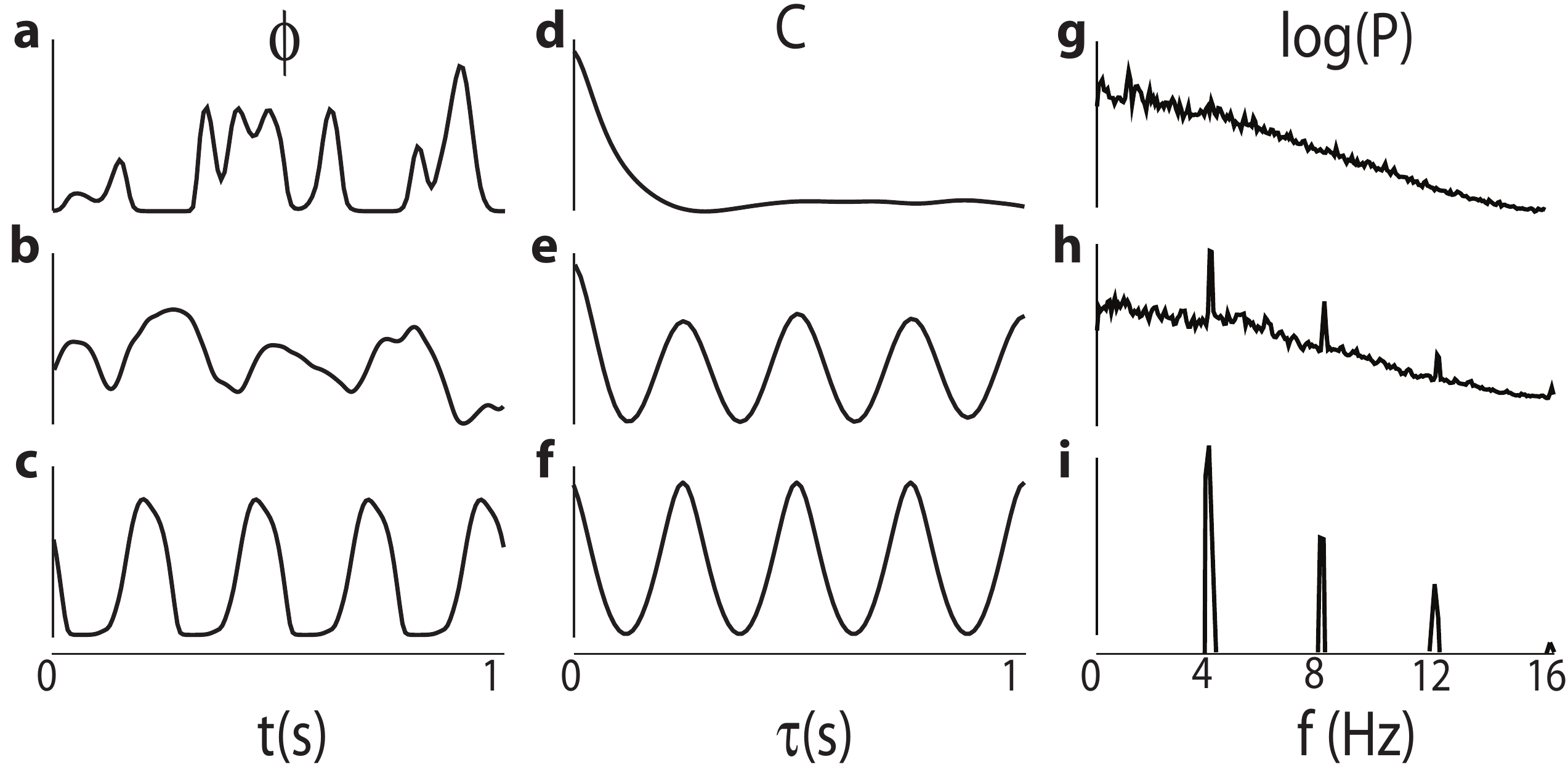}
\caption{\small{Activity of typical network units (left column), average autocorrelation function (middle column) and log-power spectrum (right column) for a network with $N\eq 1000$, $g\eq 1.5$ and $r_0\eq 0.2$. a) With no input ($I\eq 0$), network activity is chaotic.  b) In the presence of a weak input ($I\eq 0.04, f\eq\omega/2\pi\eq 4$ Hz), an oscillatory response is superposed on chaotic fluctuations.  c) For a stronger input ($I\eq 0.2, f\eq 4$ Hz), the network response is periodic. d, e, f) Average autocorrelation function and g, h, i) Logarithm of the power versus frequency for the network states corresponding to panels a, b, and c\@.}}
\end{figure} 

When this network is driven with a relatively weak sinusoidal input (Fig.\ 1b, e \& h), the single-neuron response consists of periodic activity induced by the input superposed on a chaotic background (Fig.\ 1b).  The average autocorrelation function for the network driven by weak periodic input consequently reveals a mixture of periodic and chaotic activity (Fig.\ 1e).  Periodic oscillations at the input frequency appear at large values of $\tau$, but the variance given by $C(0)$ is larger than the height of the peaks in these oscillations.  This indicates that the total firing-rate variance is not completely accounted for by the oscillatory response of the network to the external drive,  the additional variance arising from residual chaotic fluctuations.  Similarly, the power spectrum shows a continuous component generated by the residual chaos, a prominent peak at the frequency of the input, and peaks at harmonics of the input frequency arising from network nonlinearities (Fig.\ 1h).  

When the amplitude of the input is increased sufficiently, the single-neuron firing rates oscillate at the input frequency in a perfectly periodic manner (Fig.\ 1c), yielding a periodic autocorrelation function (Fig.\ 1f).  $C(0)$ now matches the height of the peaks in each of its subsequent oscillations, meaning that the periodic component in $C$ accounts for the entire response variance quantified by $C(0)$.   All of the network power is focused at the frequency of the input and its harmonics, also indicating a periodic response free of chaotic interference (Fig.\ 1i).  

To explore these results analytically and more systematically, we developed dynamic mean-field equations appropriate for large N. The mean-field theory is based on the observation that the total recurrent synaptic input onto each network neuron can be approximated as Gaussian noise $[2]$. The temporal correlation of this noise is calculated self-consistently from the average autocorrelation function of the network.  We begin by writing $x_i\eq x_i^0 + x_i^1$, where $x^0$ is the steady-state solution to $dx_i^0/dt\eq -x_i^0 + I\cos(\omega t + \theta_i)$ and $x_i^1$ satisfies $dx_i^1/dt\eq -x_i^1 + \sum_jJ_{ij}\phi(x_j^1 + x_j^0)$.  This implies that $x^0_I(t) \eq h\cos(\omega t + \tilde{\theta}_i)$, where $h \eq I/\sqrt{1 + \omega^2}$ and we have incorporated a frequency-dependent phase shift into the factor $\tilde{\theta}_i$.  Mean-field theory replaces the network interaction term in the equation for $x_i^1$ by a Gaussian random variable $\eta$, so that $dx_i^1/dt = -x_i^1 + \eta_i$.  Averages over time and network units (denoted by square brackets) as in Eq.\ \ref{CDef}, are implemented by averaging over $J$, $\theta$ and $\eta$, an approximation valid for large $N$. 

Self-consistence is obtained in the mean-field theory by requiring that the first two moments of $\eta$ match the moments of the network interaction that it represents.  Thus, we set  $\left[ \eta_i(t)\right] \eq \left[ \sum_j J_{ij}\phi(x_j(t))\right] \eq 0$, because $\left[J_{ij}\right] \eq 0$.  Similarly, using the identity $\left[J_{il}J_{jk}\right] \eq g^2\delta_{ij}\delta_{kl}/N$, we find that
\begin{eqnarray}
\left[\eta_i(t)\eta_j(t+\tau)\right] &=& \left[\sum_{l=1}^N J_{il}\sum_{k=1}^N J_{jk}\phi(x_l(t))\phi(x_k(t+\tau))\right] \nonumber \\
= \delta_{ij} \frac{g^2}{N}\sum_{k=1}^N&&\left[\phi(x_k(t))\phi(x_k(t+\tau))\right] 
= \delta_{ij}g^2C(\tau) \, .
\label{etaDef}
\end{eqnarray}
Next, defining $\Delta(\tau) = \left[x^1_i(t)x^1_i(t+\tau)\right]$ and recalling that $dx_i^1/dt = -x_i^1 + \eta_i$, it follows that
\b
\frac{d^2\Delta(\tau)}{d\tau^2} = \Delta(\tau) - g^2C(\tau) \, .
\label{DeltaEq0}
\e
The final step in the derivation of the mean-field equations is to note that because $x^1(t)$ and $x^1(t\!+\!\tau)$ are driven by Gaussian noise, they are Gaussian random variables with moments
$\left[ x^1(t)\right]\eq  \left[ x^1(t+\tau)\right]\eq 0$ , $\left[ x^1(t)x^1(t)\right]\eq \left[ x^1(t\!+\!\tau)x^1(t\!+\!\tau)\right]\eq \Delta(0)$, and $\left[ x^1(t\!+\!\tau)x^1(t)\right]\eq \Delta(\tau)$.  To realize these constraints, we introduce three Gaussian random variables with zero mean and unit variance, $z_i$ for $i\eq1, 2, 3$, and write 
\[
x^1(t) = \sqrt{\Delta(0) \!-\! |\Delta(\tau)|}z_1 + \sgn(\Delta(\tau))\sqrt{|\Delta(\tau)|}z_3
\]
and
\[
x^1(t+\tau) = \sqrt{\Delta(0) \!-\! |\Delta(\tau)|}z_2 + \Delta(\tau)\sqrt{|\Delta(\tau)|}z_3 \, .
\]
$C$ can then be computed by writing $x\eq x^0 + x^1$ and integrating over these Gaussian variables,
\begin{eqnarray}
C(\tau) &\eq&\int_{0}^{2\pi}\!\!\frac{d\theta}{2\pi}\!\int_{-\infty}^{\infty}\!\!\!\!\!Dz_3\!\int_{-\infty}^{\infty}\!\!\!\!\!Dz_1\, \phi\!\left(\sqrt{\Delta(0) \!-\! |\Delta(\tau)|}z_1 \right.\nonumber\\
&& \hspace*{0.2in} \left.\left. + \sgn(\Delta(\tau))\sqrt{|\Delta(\tau)|}z_3\!+\! h\cos\left(\theta\right)\right) \right. \nonumber \\
&\times& \int_{-\infty}^{\infty}\!\!\!\!\!Dz_2 \label{CEq}\phi\Big(\sqrt{\Delta(0) \!-\! |\Delta(\tau)|}z_2
+ \sqrt{|\Delta(\tau)|}z_3  \nonumber\\
&& \hspace*{0.2in} + h\cos\left(\omega\tau \!+\! \theta\right)\Big) \, , 
\end{eqnarray}
where $Dz_i = dz_i\exp(-z_i^2/2)/\sqrt{2\pi}\;$ for $\;i\eq 1,2,3$ and $\;\theta \eq \tilde\theta \!+\! \omega t$.  Eq.\ \ref{CEq} determines $C(\tau)$ as a nonlinear function of $\Delta(\tau)$. Substituting this expression into Eq.\ \ref{DeltaEq0} provides a nonlinear differential equation for $\Delta(\tau)$, with $g$, $h$, $\omega$ and $\Delta(0)$ as parameters. 

Eq.\ \ref{DeltaEq0} has the form of the equation of motion for a classical particle of unit mass and position $\Delta(\tau)$ moving under the influence of a force that depends on $C$. This force is a function of the current position of the particle, $\Delta(\tau)$ (as well as on its initial position $\Delta(0)$), and it contains terms representing external forcing that are periodic in $\tau$ with period $2\pi/\omega$.  For weak inputs and $g$ greater than but close to 1, Eq.\ \ref{DeltaEq0} reduces to an undamped forced Duffing oscillator, although we do not restrict our analysis to this limit. 

The analogous mechanics problem has to be solved with the initial condition $\dot\Delta(0)\eq 0$, which imposes a smoothness constraint on the correlation function. The initial value $\Delta(0)$ is fixed by requiring that $\Delta(0)\!\geq\!\Delta(\tau)$.  We solved Eq.\ \ref{DeltaEq0} numerically using iterative methods to determine $\Delta(0)$, and found two types of solutions.  The first is a solution in which $\Delta(\tau)$ is a periodic function of $\tau$ with frequency $\omega$, as in Fig.\  1f. This solution, which represents a network state that is fully entrained by the oscillatory input, exists for all values of $I$, $\omega$ and $g$. The second solution is characterized by  $\Delta(\tau)$ that decays for small $\tau$ and oscillates for large $\tau$, so that $\Delta(0)$ is larger than the peaks in the large-$\tau$ oscillations, as in Fig.\  1e.  This solution, which corresponds to a non-periodic state only partially locked to the oscillatory drive, only exists for $I$ smaller than a critical value that depends on $\omega$ and $g$.  A linear perturbation analysis of the mean field theory shows that this non-periodic solution is stable throughout the regime where it exists.  The periodic solution is unstable in this regime and is stable outside it.  The mean-field analysis also shows that the non-periodic solution corresponds to a state with ``exponential" sensitivity to initial conditions (a positive Lyapunov exponent) $[2]$, i.e., a chaotic state. 

\begin{figure}
\includegraphics[width=\textwidth]{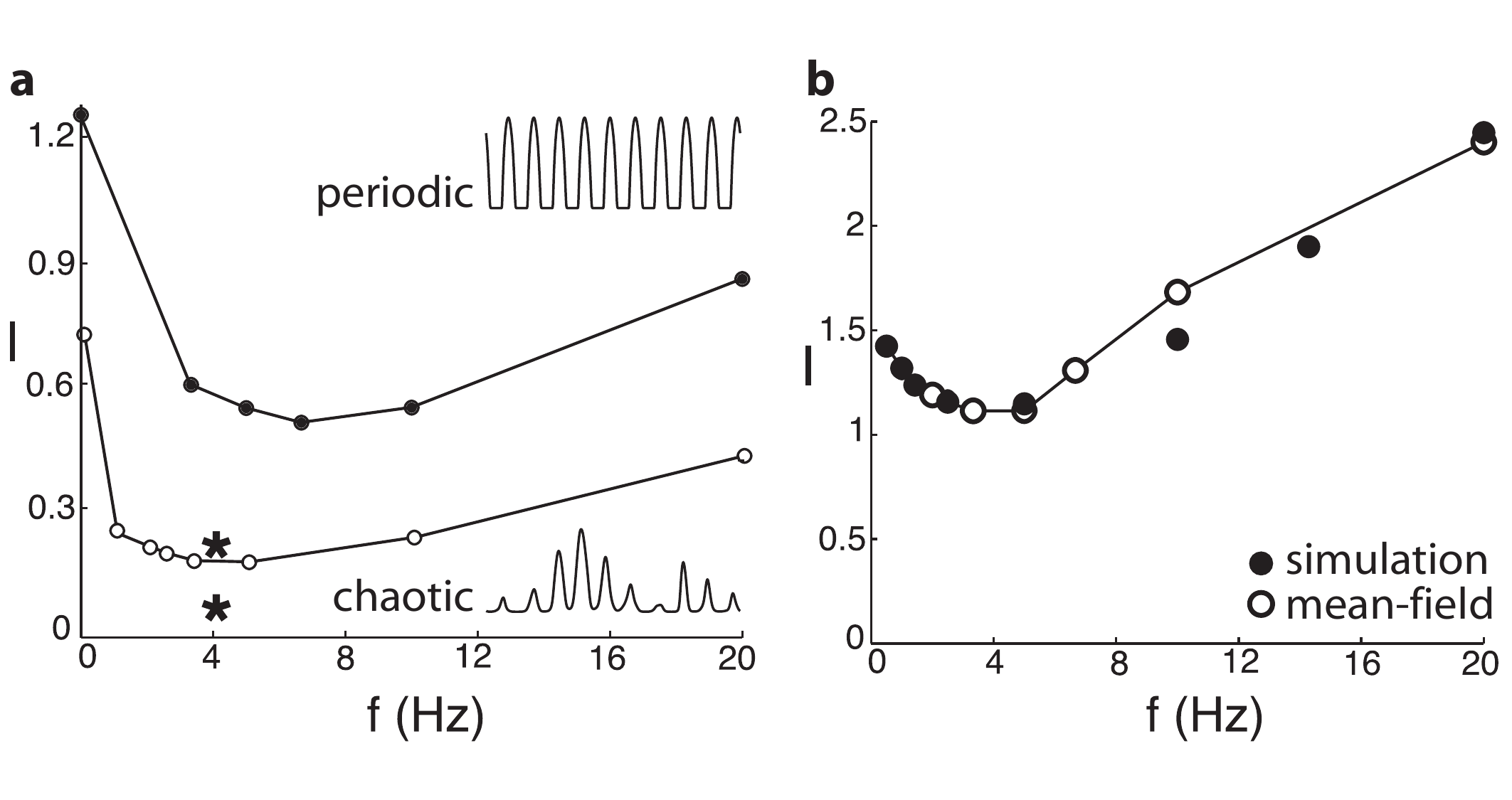}
\caption{\small{Phase transition curves showing the critical input amplitude that divides regions of periodic and chaotic activity as a function of input frequency.  a) Transition curves for $r_0\eq 0.2$ and $g\eq 1.5$ (dashed) or $g\eq 1.8$ (solid).  The stars indicate parameter values used in Figs.\  1b, e, h and 1 c, f, i\@.  The inset traces show representative single-unit firing rates for the regions indicated. b) A comparison of the transition curve computed by mean-field theory (open circles and line) and by simulating a network (filled circles) for $r_0\eq 1$, $g\eq 2$ and, for the simulation, $N\eq$ 10,000\@.}}
\end{figure}

The resulting phase diagram marks the transition between the periodic and non-periodic states (Fig.\  2).  Surprisingly, the transition curves are non-monotonic functions of frequency and reveal a ``resonant" frequency at which it is easiest to entrain the chaotic network with a periodic input (even through there is no peak in the power spectrum of the chaotic activity at this frequency).  This frequency is roughly twice the inverse time constant of the chaotic fluctuations in the spontaneous state and for $g$ not too much greater than 1, the corresponding period can be an order of magnitude longer than the single-neuron time constant.  Figs.\ 2 \& 3b indicate that internally generated fluctuations are most easily suppressed by stimuli oscillating in the few Hz range. 

The phase transition curve shifts upward and to the right as $g$ increases (Fig.\  2a \& b), indicating a higher resonant frequency as well as a larger critical input amplitude.  This occurs because the chaotic activity for larger $g$ has a higher amplitude, making it more difficult to suppress, and a smaller inverse correlation time, leading to a higher resonance frequency.  The location of the phase transition computed by mean-field theory is in good agreement with simulation results for large networks (Fig.\  2b).    

To study the implications of the phase transition further, we divide network responses into signal and noise components by separating the full response variance into two terms, $\sigma^2_{\osc}$ and $\sigma^2_{\chaos}$.  For this purpose, we subtract the square of the average value of $\phi$ from $C(\tau)$ and consider the mean-subtracted correlation function, $C(\tau) - [\phi]^2$.  The signal amplitude, $\sigma_{\osc}$, is the square root of the amplitude of the oscillatory part of this correlation function for large $\tau$ (Fig.\ 3a).  The noise amplitude, $\sigma_{\chaos}$, is the square root of the difference between the value of the mean-subtracted correlation function at $\tau\eq 0$ and the peak of its oscillations (Fig.\ 3a).   In the frequency domain, $\sigma^2_{\osc}$ measures the total power in the network activity at the input frequency and its harmonics, whereas $\sigma^2_{\chaos}$ measures the residual power. 

\begin{figure}
\includegraphics[width=\textwidth]{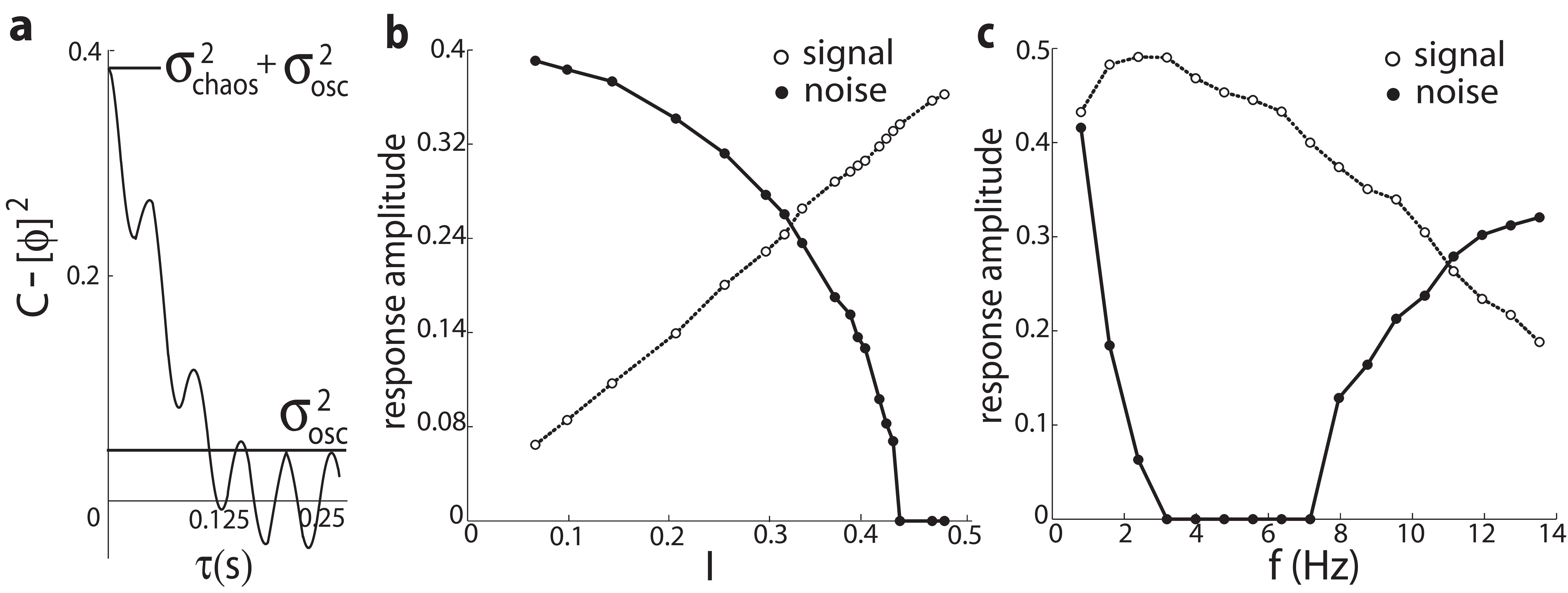}
\caption{\small{Signal and noise amplitudes as a function of input amplitude and frequency. a) Definition of the signal and noise amplitudes, $\sigma^2_{\osc}$ and $\sigma_{\chaos}$ respectively, in terms of the mean-subtracted correlation function. b) Signal and noise amplitudes for $f\eq 20$ Hz, $g\eq 1.5$ and $r_0\eq 0.2$ as a function of input amplitude.  The transition from chaotic to non-chaotic regimes occurs at $I\eq 0.44$.  c) Same as panel b, but with fixed input amplitude ($I\eq 0.2$) and varying input frequency.  In the region between 3 and 7 Hz,  responses of the network are free from chaotic noise.  In b and c, open circles denote the signal amplitude and filled circles the noise amplitude.}}
\end{figure}

The signal amplitude increases linearly with the strength of the input ($I$) over the range considered in Fig.\  3b. The noise amplitude has a more complex nonlinear dependence, reflecting the presence of the phase transition in Fig.\  2 and duplicating the effect seen in Fig.\  1, in which a sufficiently strong input completely suppresses the chaotic component of the response.   An interesting feature to note is that there is no clear signature of this chaotic-to-periodic transition in the signal amplitude.  When plotted as a function of input frequency for fixed $I$, the signal amplitude shows relatively weak frequency dependence below about 4 Hz and then rolls off at higher frequencies (Fig.\  3c).  This is a result of the low-pass filtering property of the network.  The noise amplitude has a more interesting dependence.  Between 0 and 3 Hz, the noise amplitude drops steeply and vanishes for frequencies between 3 and 7 Hz, rising again above 7 Hz.  This double transition is a consequence of the non-monotonicity of the phase transition curves in Fig.\  2. As in Fig.\  3b, there is no apparent indication of these transitions in the signal amplitude.  

It has previously been noted that chaotic activity in neuronal networks can be suppressed by either white-noise $[13]$ or constant $[14]$ input in discrete-time models.  However, discrete-time versions fail to capture the rich dynamics of the chaotic fluctuations and their effect on responses to time-dependent inputs.  Suppression of spatiotemporal chaos by periodic forcing has also been reported $[10,11,12]$, mostly through numerical simulations.  In some of these simulations, an optimal frequency for complete locking similar to Fig.\  2 has been observed $[10]$.  Our results show that such a resonance effect occurs even when the power spectrum of the unforced chaotic fluctuations falls monotonically from zero frequency (Fig.\ 1).  The networks we considered only describe the effects of fluctuations induced by local interactions, whereas additional sources of variability carried by long-range connections or by local sources of stochasticity are present in real neurons.  Therefore, we predict that an experimental plot of response variability versus stimulus frequency will follow a non-zero U-shaped curve with a minimum in the several Hz range, rather than falling to zero as in Fig.\ 3c. 

Variability in cortical responses is sometimes described by adding stochastic noise linearly to a deterministic response $[17,18]$. Our results indicate that the interaction between intrinsically generated ``noise" and responses to external drive is highly nonlinear. Near the onset of chaos, complete noise suppression can be achieved with relatively low amplitude inputs, weaker, for example, than the strength of the internal feedback.  Thus, suppression of spontaneously generated ``noise" in neural networks does not require stimuli so strong that they simple overwhelm fluctuations through saturation.  A number of experiments indicate that stimuli as well as attention can suppress firing-rate variability $[19,20,21,22,23]$(but see $[24]$).  Although other mechanisms for nonlinear suppression of neuronal variability have been proposed $[25,26,27,28,29,30]$, our analysis indicates that such suppression is a general property of the interaction between internally generated dynamics and external drive in a nonlinear network.

Spontaneous fluctuations in neural activity occur across a wide range of timescales, with increasing variability over long time intervals $[31]$ and increasing power at low frequencies, although resonances may appear $[24,32]$.  In this work we have focused on firing-rate fluctuations using smooth rate-based dynamics, not spiking dynamics.  Spiking neuron models with strong 'balanced' interactions can exhibit chaotic firing patterns $[2,3]$, but the fluctuations they produce have relatively flat power spectra associated with variability in short interspike intervals. It will be interesting to study stimulus effects in spiking network models that exhibit slow irregular modulations of firing rates. 

In our model, weak correlations (of the order of $1/\sqrt{N}$) in activity fluctuations exist between all pairs of neurons. These correlations are distributed evenly between negative and positive values across the population.  Slow spontaneous rate fluctuations in the cortex are often associated with long-range spatial correlations, especially in anesthetized animals $[33, 34]$.  As in our model, the observed spatial correlations are weaker than the firing rate autocorrelations.  In some cases, both negative and positive rate fluctuations are also observed, such that the mean value of the pairwise correlations across a populations is much smaller than the width of the distribution of correlations $[35,36,37]$. However, the extent of the contribution of local network dynamics to the observed low frequency correlations is unclear $[22,34]$.

Neuronal selectivity to stimulus features is typically studied by determining how the mean response across experimental trials depends on various stimulus parameters.  The presence of nonlinear interactions between stimulus-evoked and spontaneous fluctuating activity indicates that response components that are not locked to the temporal modulation of the stimulus may also be sensitive to stimulus parameters.  In general, our results suggest that experiments studying the stimulus-dependence of the noise component of neural responses could provide important insights into the nature and origin of activity fluctuations in neuronal circuits, as well as their role in neuronal information processing.

\sec{Acknowledgments}

KR and LA supported by National Science Foundation grant IBN-0235463 and an NIH Director's Pioneer Award (5-DP1-OD114-02), part of the NIH Roadmap for Medical Research. HS supported by grants from the Israel Science Foundation and McDonnell Foundation.  This research was also supported by the Swartz Foundation through the Swartz Centers at Columbia and Harvard.  KR's current address is Carl Icahn Laboratories, Lewis Sigler Institute for Integrative Genomics, Princeton University, Princeton NJ\@. 

\newpage
\sec{References}
\begin{list}{}{
\setlength{\leftmargin}{0.2in}
\setlength{\parsep}{\parskip}
\addtolength{\parsep}{-8pt}
\setlength{\listparindent}{-0.2in}}
\item \mbox{ }\vspace{-0.35in}

1) W.R. Softky and C. Koch,
%Cortical cells should spike regularly but do not. 
Neural Comput.\ {\bf 4}, 643Ü646 (1992).\\

2) H. Sompolinsky, A. Crisanti and H.J. Sommers,
%Chaos in Random Neural Networks.\ {\it 
Phys.\ Rev.\ Lett.\ {\bf 61}, 259-262 (1988).\\

3) C. van Vreeswijk C and H. Sompolinsky, 
%Chaos in neuronal networks with balanced excitatory and inhibitory activity 
Science {\bf 274}, 1724-1726 (1996).\\

4) N. Brunel, 
%Dynamics of networks of randomly connected excitatory and inhibitory spiking neurons. 
J.\ Physiol.\ Paris {\bf 94}, 445-463 (2000).\\

5) M. Franz and M. Zhang, Phys.\ Rev.\ E {\bf 52}, 3558-3565 (1995). \\
%Suppression and creation of chaos in a periodically forced Lorenz system.

6) I.Z. Kiss and J.L. Hudson, Phys.\ Rev.\ E {\bf 64}, 046215 (2001).\\
%Phase synchronization and suppression of chaos through intermittency in forcing of an electrochemical oscillations.

7) A.S. Pikovsky, M.G. Rosenblum, G.V. Osipov, and J. Kurths, Physica D: Nonlinear Phenomena {\bf 104}, 219-238 (1997).\\
%Issues 3-4, 1 June 1997, Pages  
%Phase synchronization of chaotic oscillators by external driving.

8) R. Brown and L. Kocarev, Chaos {\bf 10}, 344-349 (2000).\\
%doi:10.1063/1.166500 : 
%A unifying definition of synchronization for dynamical systems.

9) E.Sch{\" o}ll and H.G. Schuster (Eds), Handbook of Chaos Control, Wiley-VCH (2007).\\
%; 2nd Revision, Enlarged edition (2007) Weinheim.

10) H. Sakaguchi and T. Fujimoto,
%Elimination of spiral chaos by periodic force for the Aliev-Panfilov model.
Phys.\ Rev.\ {\bf E 67}, 067202-1:3 (2003).\\

11) A.T. Stamp, G.V. Osipov and J.J. Collins, 
%Suppressing arrhythmias in cardiac models using overdrive pacing and calcium channel blockers.
Chaos {\bf 12}, 931-940 (2002).\\

12) S. Wu, K. He and Z. Huang,
%Controlling spatio-temporal chaos via small external forces
Phys.\  Lett.\ {\bf A 260}, 345-351 (1999).\\

13) L. Molgedey, J. Schuchhardt and H.G. Schuster, 
%Suppressing chaos in neural networks by noise.\ {\it 
Phys.\ Rev.\ Lett.\ {\bf 69}, 3717-3719 (1992).\\

14) N. Bertchinger and T. Natschl\"{a}ger,
%Real-time computation at the edge of chaos in recurrent neural networks.\ {\it 
Neural Comput.\ {\bf 16}, 1413-1436 (2004).\\

15)The $\tanh$ function has the disadvantage of having the ``resting" rate $\phi(0)$ halfway between the minimum and maximum rates.  This generalization allows us to adjust the value of $\phi(0)$ to be closer to the minimum of this range, while retaining the desirable feature that the maximum of the derivative of $\phi$ is at $x\eq 0$.\\

16) The connectivity pattern in our model does not obey the restriction of cortical neurons to excitatory and inhibitory subtypes (see K. Rajan and L.F. Abbott, Phys.\ Rev.\ Lett.\ {\bf 97}, 188104 (2006) for a theoretical treatment of this problem in the linear regime).  More theoretical work is needed to establish a detailed account of the nonlinear interactions between stimulus features and ongoing fluctuations in such networks. \\

17) A. Arieli, A. Sterkin, A. Grinvald and A. Aertsen, 
%Dynamics of ongoing activity: explanation of the large variability in evoked cortical responses.\ {\it 
Science {\bf 273}, 1868-1871 (1996).\\

18) J.S. Anderson, I. Lampl, D.C. Gillespie and D. Ferster,
%Membrane potential and conductance changes underlying length tuning of cells in cat primary visual cortex.\ {\it
J.\ Neurosci.\ {\bf 21}, 2104-2112 (2001).\\

19) G. Werner and V.B. Mountcastle, 
%The Variability Of Central Neural Activity In A Sensory System, And Its Implications For The Central Reflection Of Sensory Events.
J.\ Neurophysiol.\ {\bf  26}, 958-977 (1963).\\

20) M.M. Churchland, B.M. Yu, S.I. Ryu, G. Santhanam and K.V. Shenoy,
%Neural variability in premotor cortex provides a signature of motor preparation. 
J.\ Neurosci.\ {\bf 26}, 3697-3712 (2006).\\

21) I.M. Finn, N.J. Priebe and D. Ferster, 
%The emergence of contrast-invariant orientation tuning in simple cells of cat visual cortex.
Neuron {\bf 54}, 137-152 (2007).\\

22) J.F. Mitchell, K.A. Sundberg and J.J. Reynolds, 
%Differential attention-dependent response modulation across cell classes in macaque visual area V4. 
Neuron {\bf 55}, 131-41 (2007).\\

23) M.M. Churchland et al.,\ Nature Neurosci. {\bf 13}, 369-378 (2010).\\
%B.M. Yu, J.P. Cunningham, L.P. Sugrue, M.R. Cohen, G.S. Corrado, W.T. Newsome, A.M. Clark, P. Hosseini, B.B. Scott, D.C. Bradley, M.A. Smith, A. Kohn, J.A. Movshon,  K.M. Armstrong, T. Moore, S.W. Chang, L.H. Snyder, N.J. Priebe, I.M. Finn, D. Ferster, S.I. Ryu, G. Santhanam, M. Sahani, and K.V. Shenoy,
%Stimulus onset quenches neural variability: a widespread cortical phenomenon. 

24) J.A. Henrie and R. Shapley,
%LFP Power Spectra in V1 Cortex: The Graded Effect of Stimulus Contrast
J.\ Neurophysiol. \ {\bf 94}, 479-490 (2005).\\

25) P. Kara, P. Reinagel and R.C. Reid,
%Low Response Variability in Simultaneously Recorded Retinal, Thalamic, and Cortical Neurons
Neuron {\bf 27}, 635-646 (2000).\\

26) M. Carandini,
%Amplification of trial-to-trial response variability by neurons in visual cortex.
PLoS Biol. {\bf 9}, E264 ( 2004).\\

27) P.E. Latham, B.J. Richmond, P.G. Nelson and S. Nirenberg, J.\ Neurophysiol.\ {\bf 83}, 808-827 (2000).\\

28) J. Anderson, I. Lampl, I. Reichova, M. Carandini and D. Ferster, 
%Stimulus dependence of two-state fluctuations of membrane potential in cat visual cortex. 
Nat.\ Neurosci.\ {\bf 3}, 617-621 (2000).\\

29) C.C.H. Petersen, T.T.G Hahn, M. Mehta, A. Grinvald and B. Sakmann,
%Interaction of sensory responses with spontaneous depolarization in layer 2/3 barrel cortex. 
Proc.\ Natl.\ Acad.\ Sci.\ USA {\bf 100}, 13638-13643 (2003).\\

30) B. Haider, A. Duque, A.R. Hasenstaub, Y. Yu and D.A. McCormick,
%Enhancement of visual responsiveness by spontaneous local network activity in vivo. 
J.\ Neurophysiol.\ {\bf 97}, 4186-4202 (2007).\\

31) M.V. Teich,
%Fractal character of the auditory neural spike train
IEEE Trans.\ of BioMed.\ Eng.\ {\bf 36}, 150-160 (1989).\\

32) W. Sun and Y. Dan,
% Layer-specific network oscillation and spatiotemporal receptive field in the visual cortex
Proc.\ Natl.\ Acad.\ Sci.\ (USA) {\bf 106}, 17986-17991 (2009).\\

33) M.A. Smith and A. Kohn, J.\ Neurosci.\ {\bf 28}, 12591-12603 (2008).\\

34) I. Nauhaus, L. Busse, M. Carandini and D.L. Ringach, Nature Neurosci.\ {\bf 12}, 70-76 (2008).\\

5) E. M. Maynard, N. G. Hatsopoulos, C. L. Ojakangas, B. D. Acuna, J. N. Sanes, R. A. Normann and J. P. Donoghue,
J Neurosci. {\bf 19}, 8083-8093 (1999).\\

36) A.S. Ecker, P. Berens, G.A. Keliris, M. Bethge, N.K. Logothetis and A.S. Tolias, Science {\bf 327}, 584-587 (2010).\\

37) A. Renart, J. de la Rocha, P. Bartho, L. Hollender, N. Parga, A. Reyes and K.D. Harris, Science {\bf 327} 584-590 (2010).\\

\end{list}

\end{document}